\documentclass[aps,prb,twocolumn]{revtex4} %,preprint

\usepackage{graphicx}% Include figure files
\usepackage{dcolumn}% Align table columns on decimal point
\usepackage{bm}% bold math
\usepackage{amsmath}  % subequations

\newcommand{\comment}[1]{}
\begin{document}
%\preprint{}   % Preprint number in upper right corner
\renewcommand{\theequation}{\arabic{section}.\arabic{equation}}

%\title{Transition Pressure in Superfluid Helium Capillary Flow into Vacuum}
\title{Comment on
pressure driven flow of superfluid $^4$He through a nanopipe
(Botimer  and Taborek 2016)}

%\author{}
%\thanks{}

\author{Phil Attard}
\affiliation{ {\tt phil.attard1@gmail.com}}
%\\ 20 Sept 2023}
%\noindent {\tt  Projects/QSM23/Paper5/Botimer.tex}
%\affiliation{\protect\texttt{phil.attard1@gmail.com}}

%\date{\today. Begun  20--20 Sept 2023} %phil.attard1@gmail.com
%\\
%notes in Projects/QSM23/paper2/QEoM2.tex}

\begin{abstract}
Botimer and Taborek (2016) measured the mass flux of superfluid $^4$He
through a capillary into an evacuated chamber
for various temperatures and pressures of the reservoir chamber.
They found a sharp transition from
low flux at low pressures to high flux at large pressures.
Here it is shown that the superfluid condition of chemical potential equality
predicts the induced temperature and also the transition pressure,
which is attributed to the transition from a semispherical cap
to a pool of $^4$He at the exit of the capillary.
The results show that the two-fluid equations of superfluid flow,
Landau's phonon-roton theory,
and Feynman's critical vortex theory
are unnecessary for a quantitative account
of the measured transition pressure.
\end{abstract}

%\keywords{}

\maketitle

%\newpage
%%%%%%%%%%%%%%%%%%%%%%%%%%%%%%%%%%%%%%%%%%%%%%%%%%%%%%%%%%%%%%%%%%%%%%%%%%
%
%\section{Introduction}
%\setcounter{equation}{0} \setcounter{subsubsection}{0}
%\renewcommand{\theequation}{\arabic{section}.\arabic{equation}}
%\renewcommand{\theequation}{\Alph{section}.\arabic{equation}}
%
%%%%%%%%%%%%%%%%%%%%%%%%%%%%%%%%%%%%%%%%%%%%%%%%%%%%%%%%%%%%%%%%%%%%%%%%%%

\renewcommand{\theequation}{\arabic{equation}}

\begin{figure}[t!]
\centerline{ \resizebox{8cm}{!}{ \includegraphics*{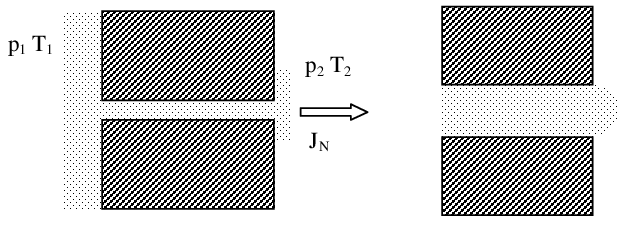} } }
% originally \Projects\QSM23\Paper2\ABC.doc
\caption{\label{Fig:Botimer-Sketch}
Liquid $^4$He in the high pressure chamber 1,
with superfluid flowing through the nanopore
and evaporating in the evacuated chamber 2.
The puddle invoked by Botimer and Taborek (2016) (left)
is compared to the semispherical cap envisaged here at low pressure drops
(magnified, right).
}
% original Projects/QSM23/Paper2/ABC.doc
\end{figure}

Botimer and Taborek (2016)
measured the mass flow rates
of superfluid $^4$He through a capillary
into an evacuated chamber
as a function of temperature and pressure of the reservoir.
Figure~\ref{Fig:Botimer-Sketch} is a sketch
of the experimental arrangement,
showing the liquid  $^4$He pooling at the exit of the capillary,
which model Botimer and Taborek (2016) invoke for their analysis.
These authors find superfluid flow below
the $\lambda$-transition temperature,
with a relatively sharp transition
from low to high flow rates with increasing reservoir pressure.
They call this the critical pressure
whereas here it is called the transition pressure.
With decreasing reservoir temperature,
the transition pressure decreases
and the relatively flat flow rates in each of the two regimes increase.

The basis of their analysis is that the flow velocity
is proportional to the square root of the chemical potential difference
(divided by mass) between the reservoir and the evacuated chamber
(Botimer and Taborek 2016, equation~(6)).
This contradicts analysis of fountain pressure measurements that show
that superfluid flow  equalizes chemical potential (Attard 2022).
Equation~(6) follows equation~(5),
%of Botimer and Taborek (2016),
%ordinally rather than logically,
which equates the acceleration of the flow
to the gradient of the kinetic energy plus the chemical potential.
But the correct conservation of momentum
hydrodynamic equation with the viscosity set to zero
has instead the gradient of the pressure (Attard 2012 equation~(5.26)).
The difference between the two involves the entropy
times the gradient of the temperature,
and so Botimer and Taborek (2016) effectively assert
that the entropy of the liquid in the capillary vanishes.
This is a highly contentious proposition for two reasons:
first the liquid in the capillary contains
a large fraction of uncondensed $^4$He,
and second the entropy of condensed bosons does not vanish
(Attard 2022, 2023a).
The application of equations~(5) and (6)
is also dubious in that the chemical potential $\mu_2$
refers to the liquid $^4$He in the pool, which is stationary.
In any case,
the classical equations of motion upon which equation~(5) is based
do not hold in the quantum condensed regime (Attard 2023b, 2023d).

%%%%%%%%%%%%%%%%%%%%%%%%%%%%%%%%%%%%%%%%%%%%%%%%%%%%%%%%%%%%%%%%%%
%%%%%%%%%%%%%%%%%%%%%%%%%%%%%%%%%%%%%%%%%%%%%%%%%%%%%%%%%%%%%%%%%%

Botimer and Taborek (2016)
calculate the temperature of the evacuated chamber
based on certain equations for the evaporation rate,
and find $T_2=$0.7--0.8\,K,
this being relatively insensitive to $T_1$
and independent of $p_1$.
This is used to calculate the chemical potential,
$\mu_2 = \mu^\mathrm{sat}(T_2)$.
The reservoir pressure at which $\mu_1 = \mu(T_1,p_1)$
first exceeds $\mu_2$ is taken by them to be
the transition pressure, $p_1 = p_\mathrm{tr}$.
Botimer and Taborek (2016) acknowledge that their calculations
predict zero flow rates below the transition pressure,
and monotonically increasing flow rates above the transition pressure,
both of which are contrary to the observed behavior.

\begin{figure}[t!]
\centerline{ \resizebox{8cm}{!}{ \includegraphics*{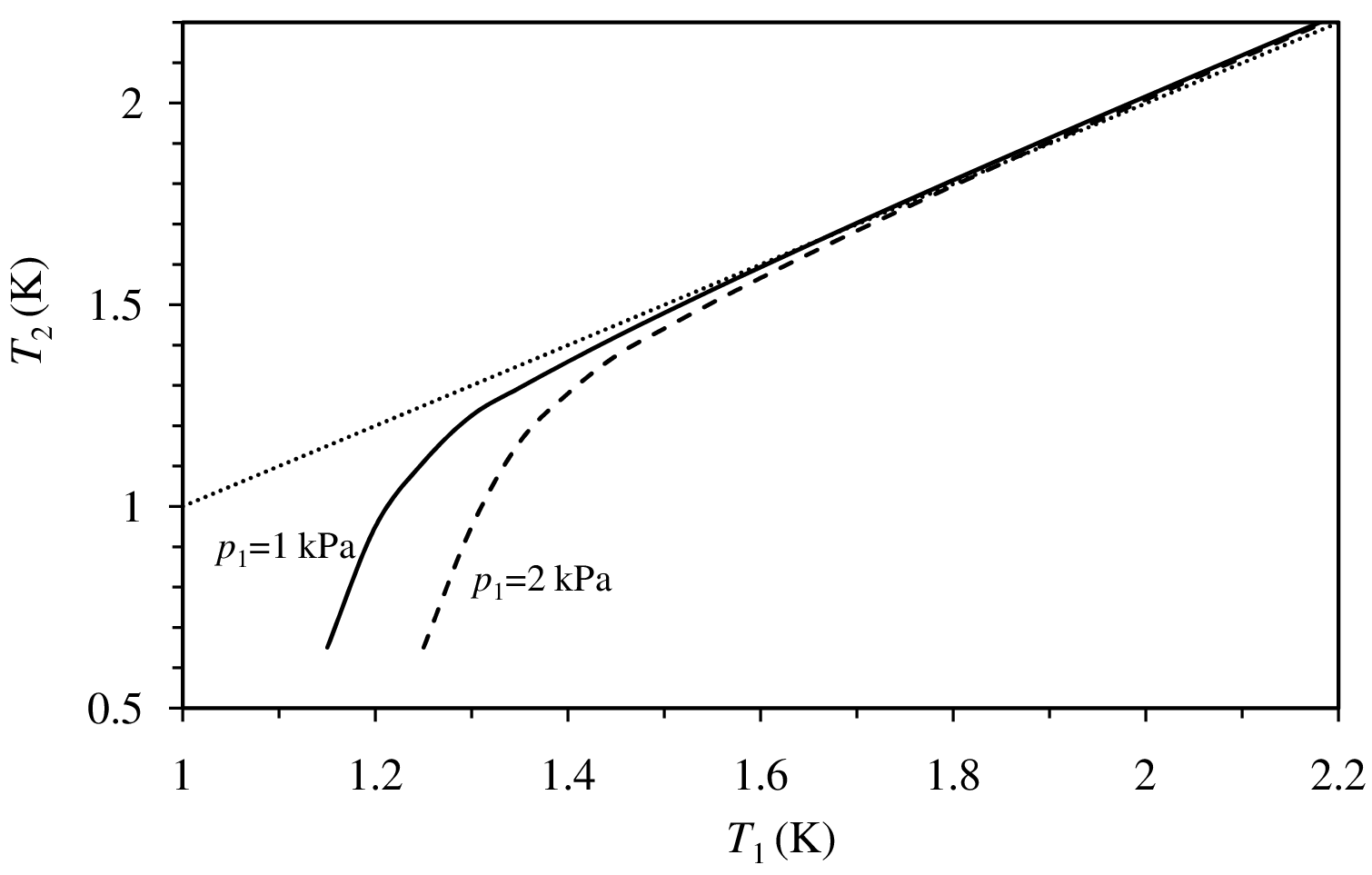} } }
\caption{\label{Fig:T2}
The temperature of the evacuated chamber $T_2$
versus the reservoir temperature $T_1$
for reservoir pressure $p_1 = 1$\,kPa (solid curve)
and $p_1 = 2$\,kPa (dashed curve),
as given by equation~(\ref{Eq:T2}).
The dotted line is $T_2=T_1$.
The thermodynamic data is taken from Donnelly and  Barenghi (1998),
with the chemical potential corrected as detailed by Attard (2022).
}
% original Projects/QSM22/Fountain.xlsx:figBot(2)
\end{figure}

As mentioned above,
the condition for superfluid flow
is that the chemical potentials are equal (Attard 2022, 2023a).
At large pressure drops the pool model advocated by Botimer and Taborek (2016)
is reasonable and $\mu_2 = \mu^\mathrm{sat}(T_2)$.
Hence the superfluid condition is
\begin{eqnarray} \label{Eq:T2}
\mu^\mathrm{sat}(T_2)
& = &
\mu(T_1,p_1)
\nonumber \\ & = &
\mu^\mathrm{sat}(T_1)
+  \rho_1^{-1} [p_1-p^\mathrm{sat}(T_1)] .
\end{eqnarray}
This determines $T_2$,
the temperature of the evacuated chamber (figure~\ref{Fig:T2}).
It can be seen that as the reservoir temperature decreases,
the evacuated chamber temperature increasingly falls
below the reservoir temperature.
The effect is larger for larger pressure drops.
%The curves cover the range of available thermodynamic data,
%which begin at $T_1=0.65$\,K,
%(Donnelly and  Barenghi (1998).

%%%%%%%%%%%%%%%%%%%%%%%%%%%%%%%%%%%%%%%%%%%%%%%%%%%%%%%%%%%%%%%%%%
%%%%%%%%%%%%%%%%%%%%%%%%%%%%%%%%%%%%%%%%%%%%%%%%%%%%%%%%%%%%%%%%%%

At small pressure drops,
the flow rate is too small to form the puddle
assumed by Botimer and Taborek (2016).
Instead it likely forms a semispherical cap
to the capillary tube (figure~\ref{Fig:Botimer-Sketch}, right).
The curvature radius of the cap is $R_\mathrm{lv}$,
with $R_\mathrm{tube} \le R_\mathrm{lv} < \infty$.
There is a Young-Laplace pressure drop across the interface,
$p_\mathrm{2,liq}-p^\mathrm{sat}(T_2)
 = 2 \gamma_\mathrm{lv}/R_\mathrm{lv}$,
where $\gamma_\mathrm{lv}$ is the liquid-vapor surface tension.
%The vapour pressure pressure in the evacuated chamber,
%$p_\mathrm{2,vap} \alt 0.07$\,Pa (Botimer and Taborek 2016),
%has been neglected.
The evaporation across this interface is arguably
proportional to the surface area of the cap,
which  is $ \pi R_\mathrm{tube}^2$ for $p_1 \rightarrow p^\mathrm{sat}(T_1)$,
and $ 2\pi R_\mathrm{tube}^2$ for $p_1 \rightarrow p_\mathrm{tr}$.
Obviously the area of the semispherical cap
is substantially less than the area of the pool that forms
at high pressure drops,
which explains the transition from low to high flow rates.
Invoking the superfluid condition $\mu_2=\mu_1$
one obtains in the semispherical cap regime
\begin{equation} \label{Eq:Rlv}
\mu^\mathrm{sat}(T_2) +\frac{ 2 \gamma_\mathrm{lv,2}}{\rho_2  R_\mathrm{lv}}
=
\mu^\mathrm{sat}(T_1)
+ \rho_1^{-1} [p_1-p^\mathrm{sat}(T_1)] .
\end{equation}
For given $T_2$, $T_1$, and $p_1$
this gives $ R_\mathrm{lv}$,
which decreases with increasing reservoir pressure.
The condition $R_\mathrm{lv} =R_\mathrm{tube}$
defines the transition pressure.

%%%%%%%%%%%%%%%%%%%%%%%%%%%%%%%%%%%%%%%%%%%%%%%%%%%%%%%%%%%%%%%%%%
\begin{figure}[b!]
\centerline{ \resizebox{8cm}{!}{ \includegraphics*{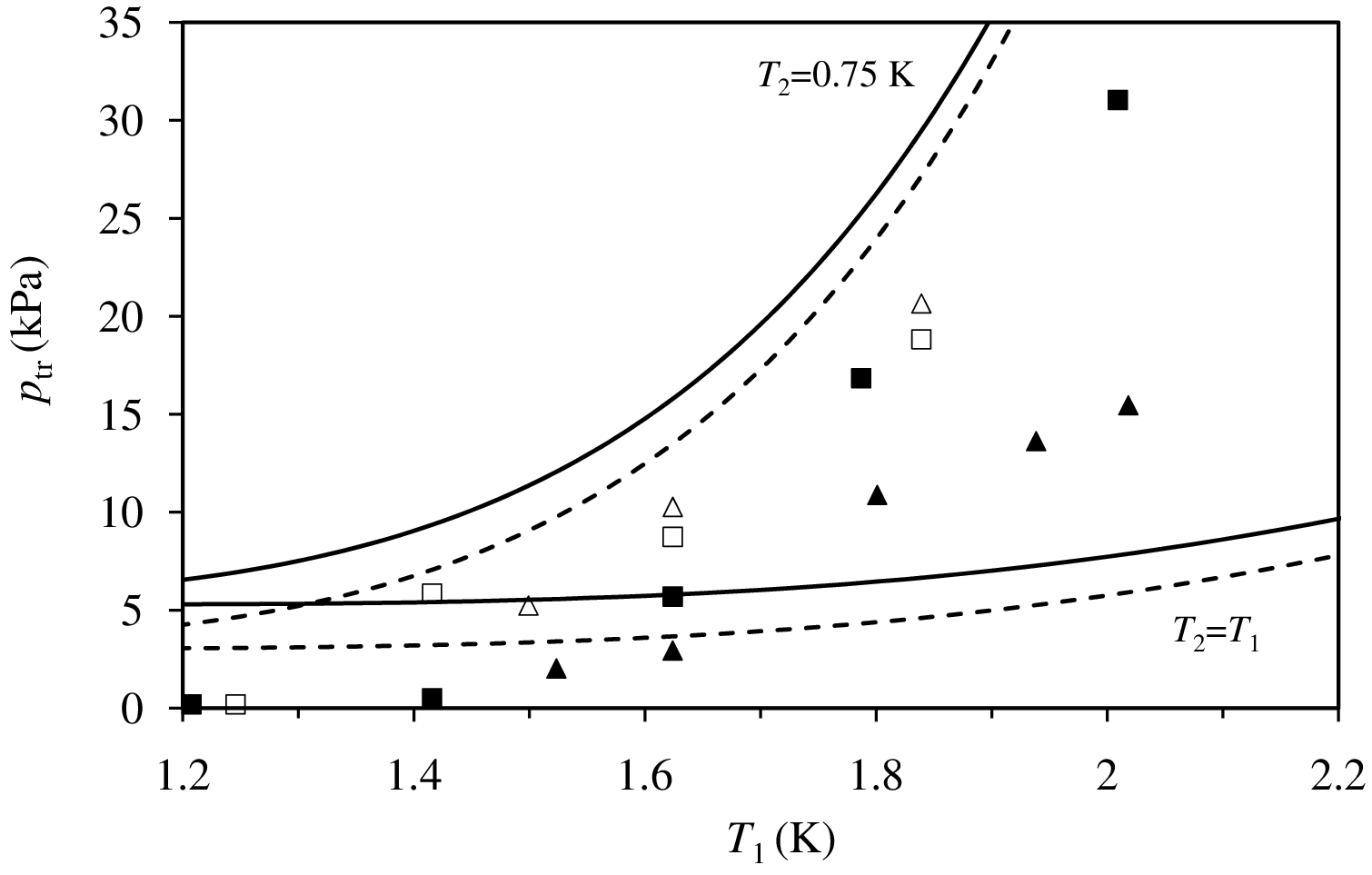} } }
\caption{\label{Fig:ptr}
Transition pressure versus reservoir temperature
for capillaries of length $L$ and radius $R_\mathrm{tube}$.
The symbols are measured data (Botimer and Taborek  2016 figure~7)
with
the filled squares being pipe~1 ($L=2$\,mm, $R_\mathrm{tube}=131.2$\,nm),
the filled triangles being pipe~2 ($L=30$\,mm, $R_\mathrm{tube}=131.7$\,nm),
the empty squares being pipe~3 ($L=1.5$\,mm, $R_\mathrm{tube}=230$\,nm),
the empty triangles being pipe~4 ($L=32$\,mm, $R_\mathrm{tube}=231.2$\,nm).
The curves are from equation~(\ref{Eq:Rlv})
with $R_\mathrm{lv} =R_\mathrm{tube}$,
with the solid curve being $R_\mathrm{tube}=131.5$\,nm
and the dashed curve being $R_\mathrm{tube}=230.5$\,nm,
and the lower pair of curves using $T_2=T_1$
and the upper pair of curves using $T_2=0.75$\,K.
The thermodynamic data is taken from Donnelly and  Barenghi (1998),
with the chemical potential corrected as detailed by Attard (2022).
}
% original Projects/QSM22/Fountain.xlsx:figBot
\end{figure}
%%%%%%%%%%%%%%%%%%%%%%%%%%%%%%%%%%%%%%%%%%%%%%%%%%%%%%%%%%%%%%%%%%

The transition pressure predicted by this
agrees semi-quantitatively
with the measured values (figure~\ref{Fig:ptr}).
The results suggest that the temperature of the second chamber
is lower than that of the reservoir,
but perhaps not quite so low as estimated by Botimer and Taborek  (2016).
%The data is not good enough to elucidate
%the variation of $T_2$ with $T_1$, $p_1$, or flow rate.

The present calculations explain
the sharp transition from low to high flow rates with increasing pressure
as being due to the transition from a semispherical cap
at low pressure drops
to, as suggested by Botimer and Taborek (2016),
a flat pool of large surface area at high pressure drops.
The present calculations do not invoke the superfluid velocity,
or attempt to predict the mass flow rates,
which Botimer and Taborek (2016) do do.

In any case,
the fact that the transition in flow rates can be accounted for
by a simple morphological change
says that dynamic considerations
such as the superfluid critical velocity,
the Feynman estimate of the critical vortex size,
or the Landau theory of rotons,
are superfluous.

%\newpage
% $\;$\\

\section*{References}

%%%%%%%%%%%%%%%%%%%%%%%%%%%%%%%%%%%%%%%%%%%%%%%%%%%%%%%%%%%%%%%%%%%%%%%%%%

\begin{list}{}{\itemindent=-0.5cm \parsep=.5mm \itemsep=.5mm}

\item %{NETDSM}
Attard  P 2012
\emph{Non-equilibrium thermodynamics and statistical mechanics:
Foundations and applications}
(Oxford: Oxford University Press)

\item  % Attard22e % QSM22/paper4
Attard P (2022)
Further On the Fountain Effect in Superfluid Helium
arXiv:2210.06666 (2022)

\item %{STD2.2}
Attard  P 2023a
\emph{Entropy beyond the second law.
Thermodynamics and statistical mechanics
for equilibrium, non-equilibrium, classical, and quantum systems}
(Bristol: IOP Publishing, 2nd edition)

\item % Attard23b % QSM23/paper
Attard  P 2023b
Quantum Stochastic Molecular Dynamics Simulations
of the Viscosity of Superfluid Helium
arXiv:2306.07538

\item % Attard23d % QSM23/paper3
Attard  P 2023d
Hamilton's Equations of Motion from Schr\"odinger's Equation
arXiv:2309.03349

\item
Botimer J and Taborek P 2016
Pressure driven flow of superfluid $^4$He through a nanopipe
\emph{Phys.\ Rev.\ Fluids} {\bf 1} 054102

\item %refd
Donnelly R J and  Barenghi C F 1998
The observed properties of liquid Helium at the saturated vapor pressure
\emph{J.\ Phys.\ Chem.\ Ref.\ Data} {\bf 27} 1217

\end{list}

%%%%%%%%%%%%%%%%%%%%%%%%%%%%%%%%%%%%%%%%%%%%%%%%%%%%%%%

\end{document}